%% file: main.tex
\documentclass[11pt]{article}
\usepackage{times}
\usepackage{geometry}
\usepackage[utf8]{inputenc}
\usepackage{enumitem,amssymb}
\usepackage{ragged2e}
\newlist{thematic}{itemize}{8}
\setlist[thematic]{label=$\square$}
\usepackage{pifont}
\usepackage{fancyhdr}
\usepackage[USenglish]{babel}
\usepackage[utf8]{inputenc}
\usepackage[T1]{fontenc}
\usepackage{newunicodechar}
\newunicodechar{−}{\textminus}

\usepackage{epsfig}
\usepackage{wrapfig}
\usepackage{color}

\usepackage[vflt]{floatflt}
\usepackage{longtable}
\usepackage{caption}
\usepackage{jheppub}   
\usepackage[dvipsnames,svgnames,x11names]{xcolor}
\usepackage[all]{hypcap} %Links go to figures; breaks on deluxetables (use \capstartfalse \capstarttrue to fix it)
\usepackage{amssymb,amsmath}
\usepackage{longtable}
\usepackage{amssymb}
\usepackage{amsmath, xspace}
\usepackage{graphics,graphicx} 
\usepackage{rotating}
\usepackage{color}
\usepackage{xcolor}
\usepackage{multirow}
\usepackage{subfigure}
\usepackage{sectsty}
\usepackage[outercaption]{sidecap}
\sidecaptionvpos{figure}{c}
\usepackage{comment}
\usepackage{biblatex}
\usepackage{enumitem}
\setlist[enumerate]{itemsep=0pt, parsep=0pt}
\setlist[itemize]{itemsep=0pt, parsep=0pt}

\begin{document}

\newcommand{\cc}[1]{{\textcolor{blue}{#1}}} %comments
\newcommand{\ignore}[1]{}

\input{macros}

\raggedright
\huge
A vision for ground-based astronomy beyond the 2030s
% how do we ensure humanity can still do ground-based astronomy next century (and beyond)?
\linebreak
\bigskip
\normalsize

\raggedright
\Large
How to build ESO's next big telescope sustainably
%How do we ensure humanity can still do ground-based astronomy in the next decades, century, and beyond?
\linebreak
\normalsize

%\cc{\begin{itemize}
%    \item ESO's guidelines and Submission details website: \href{https://next.eso.org/call-for-white-papers/}{https://next.eso.org/call-for-white-papers/}
%    \item Submission Deadline: 15 Dec 2025
%    \item ESO will not make the white papers public but the authors can post them on arXiv.
%\end{itemize}}
%\cc{As stated on the ESO Expanding Horizons Call for white papers website, they are soliciting white papers from the community (at all career stages) to encourage broad discussions across the community and help identify future challenges.}
%\cc{To help coordinate our efforts, the AtLAST team has put together this latex template for internal use, and have decided to make it available more generally for anyone who may find it useful. }
%\cc{Some of the blue text guidance in each of the suggested sections is there to help some of our more junior team members understand what to expect to need to include (and at what level). }
%\cc{Any suggestions made here reflect how we are intending to answer the call issued by ESO.}
\bigskip

\textbf{Authors:} 
%\cc{Note: first 3 authors must be from ESO countries}
%order for both ESO and ArXiv submission
\textbf{Laurane Fréour$^*$} (UNIVIE, Austria); \textbf{Mathilde Bouvier$^*$} (Leiden Observatory, The Netherlands);  \textbf{Tony Mroczkowski$^*$} (ICE-CSIC, Spain); \textbf{Callie Clontz$^*$} (MPIA, Germany); Fatemeh Zahra Majidi (INAF OACN, Italy); Vasundhara Shaw (JBCA, University of Manchester, UK); Olivier Absil (ULiège, Belgium); Anna Cabré (UPenn, USA); Olivier Lai (OCA, France); Dylan Magill (Queen's University Belfast, UK); Jake D. Turner (Cornell University, USA)

\blfootnote{*Equal contributions}
%Olivier Absil (ULiège, Belgium); \textbf{Mathilde Bouvier} (Leiden Observatory, The Netherlands), Anna Cabré (UPenn, USA); \textbf{Callie Clontz} (MPIA, Germany); \textbf{Laurane Fréour} (UNIVIE, Austria); Olivier Lai (OCA, France); Dylan Magill (Queen's University Belfast, UK); Fatemeh Zahra Majidi (INAF OACN, Italy);  \textbf{Tony Mroczkowski} (ICE-CSIC, Spain);  Vasundhara Shaw (JBCA, University of Manchester, UK); Jake D. Turner (Cornell University, USA)

\vspace{2em}
\textbf{Science Keywords:} 
sustainability, climate and earth science, habitable worlds (Earth)
%\cc{Note: THIS IS NOT A REQUIREMENT FROM ESO. We could list here a few (3-6) of the keywords ESO uses in its proposal system here to help the Senior Science Committee understand what to expect from the white paper. An overview of ESOs proposal keywords used in the recent P117 call can be found in \href{https://www.eso.org/sci/observing/phase1/p117/CfP117.pdf}{Appendix A of this document}.}
\linebreak

 \captionsetup{labelformat=empty}
\begin{figure}[h]
   \centering
\includegraphics[width=1\textwidth]{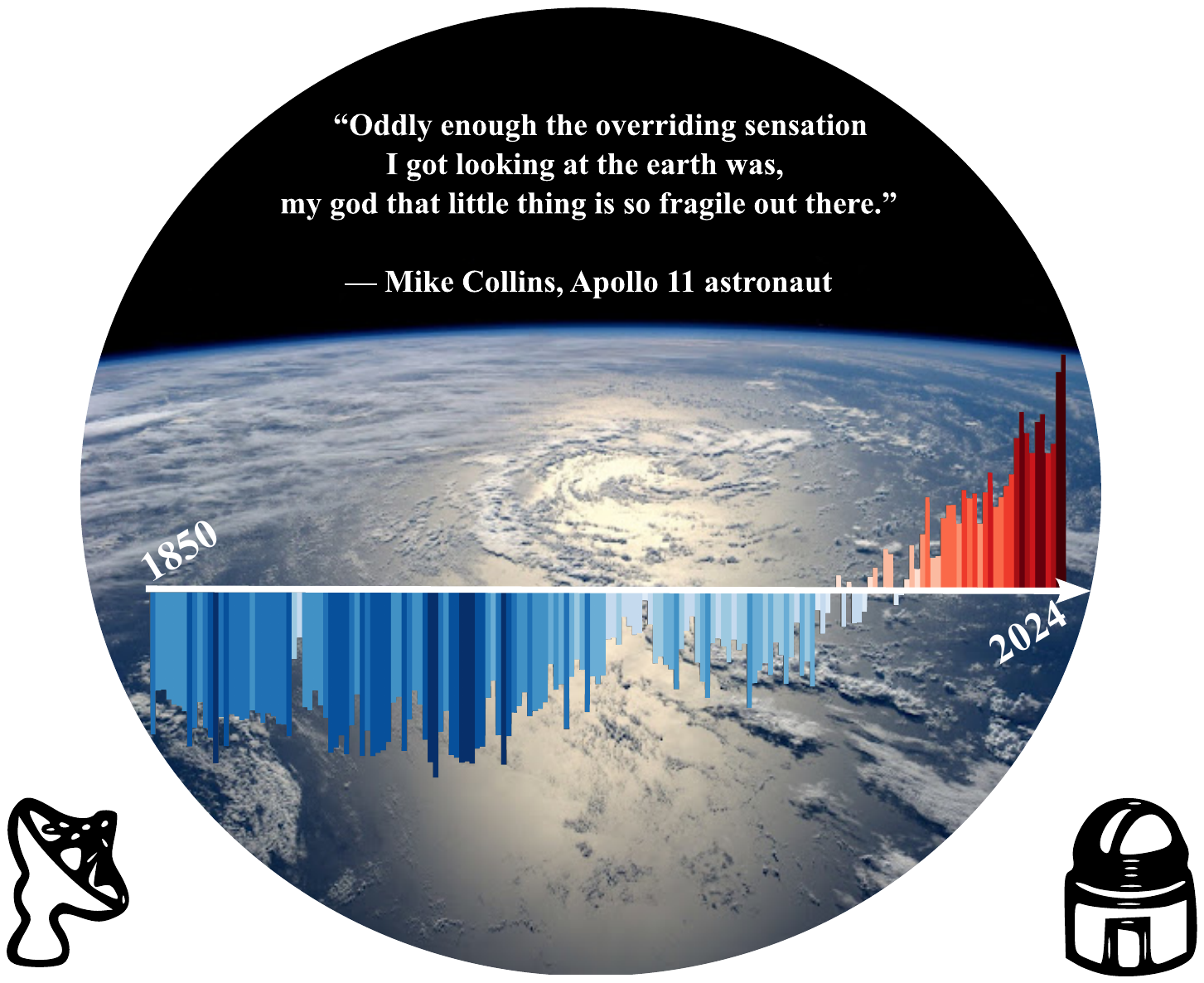}
   \caption{Earth as seen from the ISS, overlaid with global temperature anomalies (colors relative to the 1961–2010 average). Today, global temperatures have risen by 1.2$^\circ$C since 1962 -- the year ESO was founded -- based on HadCRUT5 data (\url{https://www.metoffice.gov.uk/hadobs/hadcrut5/}).
   \newline
   Images credits: adapted from NASA/ESA (Earth), \url{https://showyourstripes.info/b} (climate stripes), and from telescopes designs by Daniela Leitner (PHANGS collaboration).} 
   %\cc{I think we should have the Earth from space (maybe like \href{https://science.nasa.gov/resource/voyager-pale-blue-dot-download/}{Pale Blue Dot Revisited}?) or a map of the changes in temperature or PWV expected for the large amount of warming}
   %Carbon dioxide levels over the past 800\,000 years. Adapted from \url{climate.nasa.gov.} Telescopes images are adapted from design by Daniela Leitner.

\end{figure}
\vspace{-1mm}

% this will reset the figure counter, so the text has figure 1.
\setcounter{figure}{0}
\captionsetup{labelformat=default}

%\justify

\pagebreak
%\cc{Below are some suggested section headings for the main 3-pages long document. Modify/delete/disregard as needed to best suit your white paper.} \cc{A reminder that the request from ESO is that the white papers should include:}
%\cc{\begin{itemize}
%    \item a focus on one (or group of) science question(s) and explain why this needs a facility we do not expect to have by the 2030s
%    \item a short ($<$ half page) description of what technology developments / data handling requirements may be needed.  %not looking for a detailed description of a facility
%\end{itemize}}

\justifying

\section*{Abstract}
\vspace{-4mm}
%\cc{Suggest to keep it short, 4-5 lines}
%\textbf{Abstract:} 

%Astronomy is about understanding our place in the universe, including our home, the closest habitable world.  Astronomers attempt to think of the big picture and the long term, both over the past 13.8 billion year history of the universe and for setting a vision of the future that will inspire future generations of scientists, engineers, and others.  

 Astronomy is the study of the Universe and all the objects that it comprises. Our attention is therefore usually focused beyond Earth, home to the only form of life known today. However, how can we continue to explore the secrets of the Universe, if we stand by and watch our only home burn?  
 We know that {\it there is no Planet B}.
 %There is no planet B -- we know that. 
 It is therefore urgent that, as astronomers, we collectively work to protect the Earth, allowing future generations the opportunity to continue to uncover the secrets of the cosmos. As astronomical facilities account for the majority of our community's carbon footprint, we propose guidelines that we hold crucial for the European Southern Observatory (ESO) to consider in the context of the Expanding Horizons programme as it plans a next-generation, transformational facility.
 \vspace{-6mm}

% we provide in this white paper guidelines to help the European Southern Observatory (ESO) choose to build and operate its next transformational facility in the most sustainable way possible. 
%How do we set that example?
%How do we do astronomy in a sustainable way? 
%Can we make astronomy carbon-neutral in 2040?

\section{Context and motivation}
\vspace{-4mm}

Sustainability, defined by the United Nations Brundtland Commission (1987) as “meeting the needs of the present without compromising the ability of future generations to meet their own needs,” is under threat: of the nine planetary boundaries that define a safe operating space for humanity, seven have already been crossed (\href{https://onlinelibrary.wiley.com/doi/epdf/10.1111/gcb.70238}{Findlay et al.\ 2025}). Astronomy-related activities contribute significantly to this pressure, particularly through climate change, but also via ocean acidification, aerosol emissions, or freshwater use—highlighting the field’s responsibility to act in ways that ensure a sustainable future.

Astronomy both contributes to climate change and is exposed to its effects directly, through intensified extreme weather events such as wildfires that threaten observatories (\href{https://www.ipcc.ch/report/ar6/wg1/chapter/technical-summary/}{Arias et al.\ 2021}), and indirectly, through atmospheric changes that might impair observations (\href{https://www.nature.com/articles/s41550-020-1203-3}{Cantalloube et al.\ 2020}, \href{https://arxiv.org/pdf/2208.04918}{Haslebacher et al.\ 2022}).

Ensuring that the field can continue to create knowledge in a sustainable way is crucial to have a long-lasting positive impact on the world. Astronomy provides significant benefits to society and the economy, offering a unique perspective on our planet that can inspire action while contributing to several of the United Nations Sustainable Development Goals (SDGs) established in 2015 (\href{https://www.nature.com/articles/s41550-025-02602-x}{Mdhluli et al.\ 2025}). At the same time, astronomical activities such as stargazing can also evoke a sense of awe, which has been shown to positively impact well-being (\href{https://www.sciencedirect.com/science/article/pii/S0272494423002463}{Barnes \& Passmore, 2024}; \href{https://globalgoals.org/goals/3-good-health-and-well-being/}{SDG 3, Good Health and Well-being}). To date, astronomers have confirmed 6,045 exoplanets, yet none show signs of life, underscoring the rarity of life and the critical importance of protecting the only life-bearing planet we know: Earth.
Fortunately, a wide array of mitigation and adaptation measures within the astronomy community ensures that the field can continue to have a largely positive impact on the world.
We, the astronomical community and users of ESO facilities, have a moral obligation to adapt our practices to the science of the 21st century and address the climate crisis throughout our everyday activities (\href{https://globalgoals.org/goals/13-climate-action/}{SDG 13, Climate Action}).

This paper is not just about one astronomical result or instrument but concerns the general approach of a scientific endeavor. This is about valuing the robust scientific results from our colleagues in climate and earth sciences, using them to adapt our practices and decisions for the future in astronomy, and mitigating our environmental impact. We are writing this paper as we think ESO’s Expanding Horizons is the perfect opportunity to place sustainability at the heart of astronomy in the 2040s. We highlight below the most important points that need to be considered in terms of sustainability for the future of astronomy, and to encourage ESO to lead this change.

\vspace{-6mm}

\section{Shaping a future sustainable astronomy through observatories} 
\label{sec:obs}
\vspace{-4mm}

Observatories, both ground and space-based, are the primary source of greenhouse gas (GHG) emissions in the field of astronomy with an annual emission of about 1.2 Mt CO$_2$e/yr in 2019 (\href{https://www.nature.com/articles/s41550-022-01612-3}{Kn\"odleseder et al.\ 2022}). These research facilities are thus one of the main starting points in helping to forge a sustainable future for astronomy (\href{https://www.nature.com/articles/s41550-021-01486-x}{Burtscher et al.\ 2021}). To meet the objectives of the Paris Agreement, a recent study showed that it would be necessary to decrease the number of astronomy infrastructures by 3\% per year and double the decarbonization of existing facilities (\href{https://www.nature.com/articles/s41550-024-02346-0}{Kn\"odlseder et al.\ 2024}, \href{https://arxiv.org/pdf/2507.14510}{Kn\"odlseder \ 2025}).
Several ESO initiatives to decrease the carbon budget of its observatories have been implemented in the last decade, such as the installation of photovoltaic arrays at Paranal and La Silla. This led to a reduction of GHG emissions from electricity consumption by $\sim$50\% between 2016 and 2022 (\href{https://www.spiedigitallibrary.org/conference-proceedings-of-spie/12182/2629056/Evolution-of-electrical-power-provisioning-for-the-ESO-installations-in/10.1117/12.2629056.short}{Filippi et al.\ 2022}). Other observatories are also transitioning to greener energies. For example, the Simons Observatory plans to replace its diesel electric generators with a large-scale photovoltaic power-plant (\href{https://www.spiedigitallibrary.org/conference-proceedings-of-spie/13098/130981G/The-advanced-Simons-Observatory-green-energy-initiative--toward-an/10.1117/12.3020415.short}{Schmitt 2024}). Another initiative comes from NOIRLab, funded by the National Science Foundation (NSF) in the USA, which aims to expand the photovoltaic array on Cerro Pach\'on where Gemini South and Vera C.\ Rubin Observatory (VRO) are located. By 2027, 100\% and 60\% of the electricity consumption of the Gemini South telescope and VRO will, respectively, come from renewable sources\footnote{\url{https://noirlab.edu/public/news/noirlab2332/?lang}}. Meanwhile, the W.\ M.\ Keck Observatory has also committed to a climate action plan to reach net-zero emissions by 2035, by decarbonizing their vehicle fleet, improving efficiency, investing in greener electricity or reducing aviation footprints (\href{https://www.nature.com/articles/s41550-022-01827-4}{McCann et al.\ 2022}). It is important that new or substantially upgraded facilities do not hamper this ongoing effort. This has been the case for example with the E-ELT whose high energy demand undermined the efforts to reduce the carbon emission of the Paranal Observatory, for which the electricity consumption will be higher in 2030 compared to 2016 (\href{https://www.spiedigitallibrary.org/conference-proceedings-of-spie/12182/2629056/Evolution-of-electrical-power-provisioning-for-the-ESO-installations-in/10.1117/12.2629056.short}{Filippi et al.\ 2022}). \textit{It is therefore imperative to inscribe sustainability at the heart of the next ESO facility, right from the start.}
%it is important that the next big telescope of ESO is aligned with the ongoing efforts to reduce the carbon footprints of the observing facilities.
\vspace{-6mm}

%\section{Artificial intelligence and large data handling}
\section{Sustainable facilities and support}
\label{sec:AI}
\vspace{-4mm}

%\textcolor{red}{FZMajidi: I am adding my AI part relevant for the sustainability practices here (very much inspired by what I'm witnessing in Roman, LSST, and WST):} \\  

\noindent As astronomy moves into the 2040s, ESO is poised to develop a next-generation facility capable of transforming our understanding of the dynamic Universe. With the emergence of global time-domain infrastructures, %(LSST, SKAO, Euclid, and space-based transient missions) 
future observatories will generate massive data streams, rapid-response observing, and AI-driven analytics. The scientific returns will be extraordinary, %: real-time characterization of transients, precision spectroscopy of alert-driven events, and long-term monitoring of variable astrophysics at unprecedented depth and resolution. 
 yet the technological and computational scale needed %to support such an observatory 
raises an equally significant challenge of sustainability. %: how to ensure that this ambitious scientific vision remains compatible with ESO’s long-term sustainability commitments. \\
%\noindent A 2040s-class ESO telescope would likely integrate fast alert ingestion, near-real-time spectroscopic reduction, automated quality assessment, and extensive AI pipelines capable of analyzing millions of events per night. These systems would generate petabytes of data annually, requiring sophisticated archival strategies and continuous, high-throughput processing. However, 
Projections estimate global datacenter energy consumption at 1,000 TWh in 2030 (\href{https://www.iea-4e.org/wp-content/uploads/2025/05/Data-Centre-Energy-Use-Critical-Review-of-Models-and-Results.pdf}{Kamiya \& Coroamă 2025}), equivalent to more than one third of the EU households' energy consumption in 2022\footnote{\url{https://ec.europa.eu/eurostat/web/products-eurostat-news/w/ddn-20240605-2}}, and a total annual water withdrawal equivalent to that of half of the United Kingdom by 2027 (\href{https://arxiv.org/abs/2304.03271}{Li et al. 2025}). This poses serious challenges in terms of energy generation, cooling water requirement, and environmental impact.
With medium-scale data centers already using MW of power (equivalent to a small town), the environmental footprint and water consumption of the computational ecosystem %(datacenters, storage farms, AI training clusters, and long-term archival infrastructure) 
to support an observatory could soon exceed that of the telescope. %Data centers require enormous power input, resulting in annual energy consumption equivalent to that of entire towns [ref]. 
Storage, redundancy, cooling, and AI workloads add to this demand, and if powered by conventional energy mixes, such systems can produce thousands to tens of thousands of tons of CO$_2$ annually.    %The environmental cost of hardware manufacturing, water use for cooling, and electronic waste from frequent hardware refresh cycles further contribute to the challenge.\\
Ignoring these impacts would run counter to ESO's commitments to environmental responsibility and risk undermining the long-term viability and public trust of major scientific investments. Sustainability must therefore be integral to the telescope’s entire design and operation at all stages.%, not a post hoc consideration. 

Achieving this requires coordinated actions across energy, hardware, software, and operations. Facilities and data centers powered sustainably -- leveraging solar, wind, hydroelectric, or geothermal sources -- can drastically reduce carbon emissions. Efficiency-focused choices, using braking energy recovery (e.g.\ \href{https://ui.adsabs.harvard.edu/abs/2024SPIE13094E..0EK/abstract}{Kiselev et al.\ 2024}), low-power AI accelerators, and high-density storage solutions, can reduce overall resource consumption. Intelligent data workflows, including streaming inference, adaptive storage, and scientifically driven data triage, can limit the volume of data requiring long-term curation. Advances in cooling technology, from liquid immersion to free-air systems, offer significant reductions in energy use. And finally, efficient algorithm design, model compression, and pipeline optimization can materially decrease the computational load of routine analyses \href{https://advanced.onlinelibrary.wiley.com/doi/10.1002/advs.202100707}{(Lannelongue, Grealey, \& Inouye 2021)}.

By embedding these sustainability principles into the telescope’s conceptual design phase, ESO can ensure that its next flagship facility remains both scientifically revolutionary and environmentally responsible. \textit{A 2040s ESO telescope should not only expand the boundaries of astrophysical discovery but also serve as a model for how large international scientific infrastructures can operate within global environmental constraints.} 

\vspace{-6mm}
\section{Technical requirements and proposed guidelines}
\vspace{-4mm}
The points tackled in Sect.~\ref{sec:obs} and \ref{sec:AI} require technical considerations and anticipation continuously throughout the life of a project. Below, we suggest a few guidelines and practical recommendations.

First of all, \textbf{sustainability should be taken into account as early as possible}, for example through life cycle assessments (LCAs). 
LCAs evaluate the environmental impact of a project throughout its life cycle, from raw material extraction to end-of-life processes. An LCA is an iterative process in which material masses, energy consumption, transportation, and more are used as inputs and translated into measurable environmental impacts across various categories, including climate change, resource use, chemical toxicity, etc. 
This guiding principle is generally applicable to e.g.\ observatories, data centers, work places, etc., and we encourage ESO to minimize environmental impact and strive towards carbon neutrality in their LCAs.
Existing facilities have already started to perform such LCAs to document their environmental impacts (e.g., \href{https://www.sciencedirect.com/science/article/abs/pii/S0927650523000890#preview-section-abstract}{Vargas-Ibáñez et al.\ 2023}, \href{https://www.nature.com/articles/s41550-024-02326-4}{dos Santos Ilha et al.\ 2024}).
As alternative design and material choices (e.g.\ those produced using green energy or incorporating carbon capture) can be assessed and modified throughout a project's design and review, the results from LCAs should be taken into account from the beginning in order to be useful and informative \href{https://www.sciencedirect.com/science/article/pii/S0360544223019643}{(Viole et al.\ 2023}, \href{https://www.sciencedirect.com/science/article/pii/S0306261924007177}{2024)}. 
%As alternative design and material choices can be easily suggested and modified during the assessment to evaluate several design options, the results of LCAs should be taken into account early on to inform design choices, the costs of which can be further assessed \href{https://www.sciencedirect.com/science/article/pii/S0360544223019643}{(Viole et al.\ 2023)}. 

Second, \textbf{sustainability is multifaceted}. Broader consideration of the impacts on biodiversity and on people locally and globally (social sustainability) should be taken into account. This can be done by soliciting biodiversity experts and involving them in the site selection. Considerations on the energy generation systems could also be made in close collaboration with local communities, who are often left behind, even though the projects impact them most. Sustainable solutions can positively benefit various local stakeholders (e.g.\ \href{https://www.nature.com/articles/s41893-024-01442-3}{Valenzuela-Venegas et al.\ 2024}), and should therefore be given extra weight in the decision.

Next, \textbf{any future facility should report transparently and regularly on its full environmental impacts}, providing not only CO$_2$ emissions numbers but also all the parameters and data that led to the results (e.g., construction, energy use, water consumption, heat generation, travel and personnel support, maintenance, and component replacement; this also includes data transport, storage, analysis, and curation). These considerations underscore ESO's core values to provide clear and open communication\footnote {\url{https://www.eso.org/public/about-eso/mission-vision-values-strategy/}}, and the broader principle of open science. While disinformation on climate change explodes through social media and news, building trust requires transparent reporting to enable reproducible results. 
Applying FAIR principles, ensuring that data products are Findable, Accessible, Interoperable, and Reusable and embedding these considerations early in the design of instruments, pipelines, and archives, enhances scientific return, reduces duplication of effort, and limits unnecessary storage and processing of poorly curated data.
In the planning phase, data management and metadata plans should explicitly address each FAIR dimension and be integrated into technical requirements. During data acquisition and reduction, observatories should enforce structured file naming, robust version control, and the use of open, community-accepted formats appropriate for long-term curation in astronomy. In the publication and preservation phases, science-ready and higher-level data products should be deposited in trusted archives with rich, machine-actionable metadata, persistent identifiers, and well-documented provenance.
Within FAIR, interoperability is typically the most technically demanding aspect, but it is crucial for future facilities that are expected to operate in a highly networked, time-domain ecosystem. In practice, this primarily entails aligning data models and access protocols with established astronomical community standards, such as those developed within the International Virtual Observatory Alliance (IVOA), and providing sufficient support so that instrument teams adopt these conventions at an early stage in their workflows.

Last, but not least, the future facility should define long-term sustainability goals and actively track their progress\footnote{e.g., \url{https://open-sdg.org/}}. The long timescales involved in observatory development, the rapidly changing global context, and the accelerating global warming require a long-term vision to ensure that sustainable considerations are not left behind once the facility becomes operational. This should apply both to environmental aspects as well as social ones, through Equity, Diversity, and Inclusion policies.

\smallskip
\textbf{Building ESO's next big telescope sustainably is a substantial challenge, but so is the opportunity: to demonstrate that cutting-edge astronomy and sustainability can advance together, shaping a future in which scientific ambition and planetary stewardship are mutually reinforcing rather than mutually exclusive.}

\ignore{
{\bf References:}
\vspace{-3mm}
\begin{itemize}
\item Energy recovery system for large telescopes: \href{https://ui.adsabs.harvard.edu/abs/2024SPIE13094E..0EK/abstract}{Kiselev et al.\ (2024)};
\item A renewable power system for an off-grid sustainable telescope fueled by solar power, batteries and green hydrogen: 
\href{https://www.sciencedirect.com/science/article/pii/S0360544223019643}{Viole et al.\ (2023)};
%\item Trade-offs between CO2 emissions and mineral resource depletion/water use of different renewable energy system scenarios: \href{https://link.springer.com/article/10.1007/s11367-024-02288-9}{Viole et al.\ (2024a)};
\item Integrated life cycle assessment in off-grid energy system design: \href{https://www.sciencedirect.com/science/article/pii/S0306261924007177}{Viole et al.\ (2024)};
\item A renewable and socially accepted energy system for astronomical telescopes:
\href{https://www.nature.com/articles/s41893-024-01442-3}{Valenzuela-Venegas et al.\ (2024)};
% papers suggested in Brainstorming Canvas -------------------------
%\item Astronomy’s carbon footprint is sky-high: \href{https://www.nature.com/articles/d41586-022-00824-x}{Research Highlight (2022)};
\item Estimate of the carbon footprint of astronomical research infrastructures: \href{https://doi.org/10.1038/s41550-022-01612-3}{Knödlseder et al. (2022)};
% links suggested in Brainstorming Canvas -------------------------
\item Open Source Data in support of the UN Sustainable Development Goals (SDG) \href{https://open-sdg.org/}{https://open-sdg.org/}
\item Findable, Accessible, Interoperable, Re-purposable (FAIR) data infrastructures \\ \href{https://www.nature.com/articles/sdata201618}{https://www.nature.com/articles/sdata201618}

% papers suggested by Sarah
\item \href{https://ui.adsabs.harvard.edu/abs/2025NatAs...9..936M/abstract}{Astronomy as a strategic driver for sustainable development}

\item \href{https://ui.adsabs.harvard.edu/abs/2024NatAs...8.1478K/abstract}{Scenarios of future annual carbon footprints of astronomical research infrastructures} \cc{This should definitely be included! One of the figures could also be adopted here. I think that would be more informative than another pale blue dot or earthrise. }

\item \href{https://ui.adsabs.harvard.edu/abs/2022NatAs...6.1223M/abstract}{A path to net-zero carbon emissions at the W. M. Keck Observatory}

\item \href{https://ui.adsabs.harvard.edu/abs/2020NatAs...4..843S/abstract}{The imperative to reduce carbon emissions in astronomy}

\item \href{https://ui.adsabs.harvard.edu/abs/2021NatAs...5..857B/abstract}{Forging a sustainable future for astronomy}

\item \href{https://ui.adsabs.harvard.edu/abs/2020NatCo..11..233V/abstract}{The role of artificial intelligence in achieving the Sustainable Development Goals}

\item \href{https://ui.adsabs.harvard.edu/abs/2020NatAs...4..819P/abstract}{The ecological impact of high-performance computing in astrophysics}

\end{itemize}
}

% for proper bibtex references; 
%there's got to be a better way to do this... maybe \nocite{*}?
% \bibliographystyle{unsrturl}
%\bibliographystyle{unsrturltrunc6}
%\bibliography{pubslist}
\nocite{*}
\printbibliography

\end{document}

%% file: macros.tex
%--- VARIOUS
\newcommand{\etal}{~et~al. }                    
\newcommand{\ojo}{\fbox{\bf !`$\odot$j$\odot$!}} 
\newcommand{\lai}{\'{\i}}
\newcommand{\y}{{\it y\ }}
\newcommand{\titlebar}{\rule{15cm}{1mm}\\[2.pt]\rule{15cm}{0.4mm}\\} 
\newcommand{\eg}{$e.g.$}
\newcommand{\ie}{{\it i.e.}}

%--- UNITS
\newcommand{\arcmin}{\hbox{$^\prime$}}               % arcmin
\newcommand{\arcsec}{\hbox{$^{\prime\prime}$}}       % arcsec
\newcommand{\kms}{ km\ s$^{-1}$}                     % km s-1
\newcommand{\ergs}{erg s$^{-1}$}                     % erg s-1
\newcommand{\ergsM}{erg s$^{-1}$Mpc$^{-3}$}          % erg s-1 Mpc-3
\newcommand{\sen}{mJy/$\sqrt{\rm Hz}$}               % Unidades de sensibilidad
\newcommand{\micron}{\rm \mu m}                            % Micras!!!
\newcommand{\cm}{{\rm cm}}
\newcommand{\mm}{{\rm mm}}

\definecolor{DarkGreen}{rgb}{0.0, 0.3, 0.0}
\definecolor{purple}{rgb}{0.5, 0.0, 0.5}
\definecolor{red}{rgb}{1, 0.0, 0.0}
\definecolor{green}{rgb}{0, 1.0, 0.0}

%% Definitions of useful commands
\newcommand{\Yx}{\mbox{$Y_{\mbox{\tiny X}}$}}
\newcommand{\Tx}{\mbox{$T_{\mbox{\tiny X}}$}}
\newcommand{\Te}{\mbox{$T_{e}$}}
\newcommand{\Isze}{\mbox{$I_{\mbox{\tiny SZE}}$}}
\newcommand{\sigT}{\mbox{$\sigma_{\mbox{\tiny T}}$}}
\newcommand{\Tcmb}{\mbox{$T_{\mbox{\tiny CMB}}$}}
\newcommand{\kB}{\mbox{$k_{\mbox{\tiny B}}$}}
\newcommand{\nH}{\mbox{$n_{\mbox{\tiny H}}$}}
\newcommand{\NH}{\mbox{$N_{\mbox{\tiny H}}$}}
\newcommand{\LameH}{\mbox{$\Lambda_{e \mbox{\tiny H}}$}}
\newcommand{\Lamee}{\mbox{$\Lambda_{ee}$}}
\newcommand{\rhogas}{\mbox{$\rho_{\mbox{\scriptsize gas}}$}}
\newcommand{\Mgas}{\mbox{$M_{\mbox{\scriptsize gas}}$}}
\newcommand{\Mtot}{\mbox{$M_{\mbox{\scriptsize tot}}$}}
\newcommand{\Yint}{\mbox{$Y_{\mbox{\scriptsize int}}$}}
\newcommand{\Ycyl}{\mbox{$Y_{\mbox{\scriptsize cyl}}$}}
\newcommand{\Ysph}{\mbox{$Y_{\mbox{\scriptsize sph}}$}}
\newcommand{\fgas}{\mbox{$f_{\mbox{\scriptsize gas}}$}}
\newcommand{\LCDM}{\mbox{$\Lambda$CDM}}
\newcommand{\h}{\mbox{$^{\mbox{h}}$}}
\newcommand{\m}{\mbox{$^{\mbox{m}}$}}
\newcommand{\s}{\mbox{$^{\mbox{s}}$}}
\newcommand{\muJ} {\mbox{{$\mu\mbox{Jy}$} beam$^{-1}$}}
\newcommand{\Omegabeam}{\mbox{$\Omega_{\mbox{\tiny beam}}$}}
\newcommand{\expf}[1]{{{\rm e}^{#1}}}

 \newcommand\blfootnote[1]{%
  \begingroup
  \renewcommand\thefootnote{}\footnote{#1}%
  \addtocounter{footnote}{-1}%
  \endgroup
}
\newcommand{\rHz}{$\rm\mu\mbox{K}\,\mbox{s}^{1/2}$} 
\newcommand{\JyHz}{$\rm mJy\,s^{\frac{1}{2}}$}
\newcommand{\uJy}{\,{\rm \mu Jy} }

\newcommand{\op}[1]{\mbox{$\tau_{\mbox{\tiny #1~GHz}} $}} 
\newcommand{\HII}{$\mathrm H\:II$} 
\newcommand{\et}{{\it et al. }}

\newcommand{\ergseq}{{\rm erg}\: {\rm s}^{-1}\:}     % erg s-1  (eqn)
\newcommand{\Hub}{km\ s$^{-1}$ Mpc$^{-1}$}           % km s-1 Mpc-1

%Solar:
\newcommand{\Msun}{$M_{\odot}$}                       % M_o 
\newcommand{\Lsun}{$L_{\odot}$}                       % L_o 
\newcommand{\MJ}{$M_J$}                                         % M_Jup
\newcommand{\ME}{$M_{\oplus}$}                          % M_Earth
\newcommand{\Moon}{$M_{Moon}$}                          % M_Moon

%Cosmology:
\newcommand{\Om}{$\Omega_{\rm M}$}
\newcommand{\Ol}{$\Omega_\Lambda$}
\newcommand{\Ok}{$\Omega_k$}

%----------------------- BIBTEX --------------------------
%--- JOURNALS
\newcommand{\apj}{\rm ApJ}
\newcommand{\apjl}{\rm ApJL}
\newcommand{\apjs}{\rm ApJS}
\newcommand{\aaps}{\rm A$\&$AS}
\newcommand{\aap}{\rm A$\&$A}
\newcommand{\aapr}{\rm A$\&$AR}
\newcommand{\mnras}{\rm MNRAS}
\newcommand{\jcap}{\rm JCAP}
\newcommand{\aj}{\rm Astron. J.}
\newcommand{\araa}{\rm ARAA}
\newcommand{\nat}{\rm Nature}
\newcommand{\pasj}{\rm PASJ}
\newcommand{\ASP}{\rm ASP COnference Series}
\newcommand{\CASP}{\rm Comm. Astrophys. Space Phys.}
\newcommand{\astroph}{\rm astro-ph/}
\newcommand{\apss}{\rm Ap\&SS}%          % Astrophysics and Space Science
\newcommand{\qjras}{\rm QJRAS}%          % Quarterly Journal of the RAS
\newcommand{\pasp}{\rm PASP}%          % Publications of the ASP
\newcommand{\physrep}{\rm Phys. Rep.}
\newcommand{\prd}{\rm Phys/ Rev. D}
\newcommand{\procspie}{\rm Proc.\ SPIE}
\newcommand{\ssr}{Space Sci.\ Rev.}

\def\aplett{{\it Astrophys.~Lett.}} % Astrophysics Letters
\def\ao{{\it Appl.~Opt.}}           % Applied Optics

\newcommand{\rsquo}{-}

\let\astap=\aap
\let\apjlett=\apjl
\let\apjsupp=\apjs
\let\applopt=\ao

\def\K{{\rm K}}
\def\milliK{{\rm mK}}
\def\mK{\, {\rm \mu K}}
\def\muK{\, {\rm \mu K}}
\def\muJy{\, {\rm \mu Jy}}
\def\MJy{{\rm MJy}}
\def\Jy{\, {\rm Jy}}
\def\mJy{\, {\rm mJy}}
\def\sr{{\rm sr}}
\def\MJysr{\MJy/\sr}
\def\Mpc{{\rm Mpc}}
\def\GHz{{\rm GHz}}
\def\specint{{I_\nu}}
\def\bfzero{{\bf 0}}

\def\icm{${\rm cm}^{-1}\;$}
\def\3he{$^3{\rm He}$}
\def\arcdeg{\hbox{$^\circ$}}
\def\kusd{{K\$\phantom{0}}} %$
\def\et{{\it et al.}}
\def\COBE{{\sl COBE}}
\def\TBD{{\bf TBD}}

\def\blast{{\sl BLAST}}
\def\blastp{{\sl BLAST-pol}}
\newcommand{\boom}{{\it Boomerang}}
\newcommand{\map}{{\it MAP}}
\newcommand{\planck}{{\it Planck}}

%\input{aas_macros.sty}
%
% Hyphenations
%
\hyphenation{CMBR}
\hyphenation{bolo-meters}
\hyphenation{an-iso-tropy}

%--- BOOKS
\newcommand{\peacock}{\rm Cambridge University Press }
\newcommand{\avishai}{\rm Cambridge University Press }
\newcommand{\peebles}{\rm Princeton University Press }

%---------------------------------------------------------
\def\lsim{\mathrel{\lower2.5pt\vbox{\lineskip=0pt\baselineskip=0pt
           \hbox{$<$}\hbox{$\sim$}}}}

\def\gsim{\mathrel{\lower2.5pt\vbox{\lineskip=0pt\baselineskip=0pt
           \hbox{$>$}\hbox{$\sim$}}}}